\documentstyle[preprint,aps,eqsecnum]{revtex}
\begin{document}
\overfullrule=0pt  

\begin{titlepage}

\begin{flushright}
DFTUZ/95--10\\  
hep--th/9503194\\
\end{flushright}

 \vspace{0cm}
 
\begin{center}
{\large\bf Extended Dualization:
a method for the Bosonization of\\    
Anomalous Fermion Systems
in Arbitrary Dimension}

 \vspace{0.4cm}
 
{\bf Jos\'e~Luis~Cort\'es$^\dagger$, 
Elena~Rivas$^\dagger$ 
and Luis Vel\'azquez$^{\ast}$}   

 \vspace{0.2cm}

$^\dagger${\sl Departamento de F\'{\i}sica Te\'orica,\\ 
Universidad de Zaragoza,
50009 Zaragoza, Spain.}

$^{\ast}${\sl Departamento de Matem\'atica Aplicada,\\ 
Universidad de Zaragoza,
50015 Zaragoza, Spain.}

\vspace{0.2cm}

\end{center}
\vspace{0.2cm}
\begin{abstract}

The technique of extended dualization developed in this paper is
used to bosonize quantized fermion systems in arbitrary dimension 
$D$ in the low energy regime. In its original (minimal)
form, dualization is restricted  to models wherein it is possible
to define a dynamical quantized conserved charge.  We generalize
the usual dualization prescription to include systems with dynamical
non--conserved quantum currents.

Bosonization based on this extended dualization requires the
introduction of an additional rank $0$ (scalar) field together with
the usual antisymmetric tensor field of rank $(D-2)$.  Our generalized
dualization prescription permits one to clearly distinguish the
arbitrariness in the bosonization from the arbitrariness in the
quantization of the system.

We study the bosonization of four--fermion interactions 
with large mass in arbitrary dimension.  
First, we observe that dualization permits one to formally bosonize
these models by invoking the bosonization of the free massive Dirac
fermion and adding some extra model--dependent bosonic terms.

Secondly, we explore the potential of extended dualization by
considering the particular case of \underbar{chiral} four--fermion
interactions. Here minimal dualization is inadequate for calculating
the extra bosonic terms. We demonstrate the utility of extended
dualization by successfully completing the bosonization of
this chiral model.

Finally, we consider two examples in two dimensions which 
illuminate the utility of using extended dualization 
by showing how quantization ambiguities in a fermionic theory
propagate into the bosonized version. 
An explicit parametrization of the 
quantization ambiguities of the chiral
current in the Chiral Schwinger model is obtained. 
Similarly, for the sine--Gordon
interaction in the massive Thirring model the quantization
arbitrariness is explicitly exhibited and parametrized.

\end{abstract}

\end{titlepage}
\hfill

\section{Introduction}

Bosonization is a procedure for converting a given fermion field
theory into its bosonic equivalent.  This equivalence is to be
understood at the quantum level, as an equivalence between the
Green functions of the two quantum theories.  The importance of
bosonization is clear, since it permits one to investigate quantum
fermion systems by using bosonic techniques, which are always more
powerful and better developed.

The bosonization program started a considerable time ago, but for
many years the known procedures were very restrictive, only
applying to theories in two dimensional $(D=2)$
spacetime~\cite{classic,Coleman,Mandelstam,Witten}. More recently
there has been a lot of interest in extending bosonization to
higher dimensions~\cite{Daamgard-Marino}.

A new direction of inquiry has been recently opened by applying
dualization techniques to the purpose of
bosonization~\cite{BuQue2d,BuQue2dnon-ab,Frohlich,BuQue}.  Dualization
can be applied to arbitrary spacetime dimension $D$, and also to
arbitrary fermion field theories, provided that a conserved quantum
charge can be defined.  We will call this original prescription
``minimal dualization''.

Given a  $D$--dimensional  fermion system with a dynamical conserved
quantum current, minimal dualization guarantees the existence of
a bosonized version.  The explicit bosonic action is obtained by
integrating out an  auxiliary gauge field. The bosonic variable is
an antisymmetric tensor field of rank $D-2$, a $(D-2)$--form, denoted
$\Lambda^{(D-2)}$.

Minimal dualization rewrites the conserved quantized current density
${\cal J}^{(1)}$,  in terms of the real field $\Lambda^{(D-2)}$ as
\begin{equation}
{\cal J}^{(1)}= \ast d\Lambda^{(D-2)}.\label{jconserv}
\end{equation}   
Here $d$ is the exterior differential and $\ast$ the Hodge star
operation.

Alternatively, in explicit components, 
\begin{equation}
{\cal J}_\mu(x)= 
\epsilon_{\mu \mu_1 ... \mu_{D-1}}\partial^{\mu_1}
\Lambda^{\mu_2 ... \mu_{D-1}}(x).\label{jconservcomp}
\end{equation}

\noindent For $D=2$ minimal dualization~\cite{BuQue2d,Frohlich}
gives the same results as conventional bosonization~\cite{Abdalla}.

\bigskip

In this paper we will extend the  minimal dualization procedure to
include anomalous quantum fermion systems.  We define an anomalous
system as a system with a non--conserved quantum dynamical
current.  In that case, the current density ${\cal J}^{(1)}$ 
is not necessarily conserved.

\noindent In extended dualization, the bosonic equivalent action
depends on the previously introduced antisymmetric rank $(D-2)$
field, plus a rank $0$ real scalar field $\lambda^{(0)}$. The relation
between the current density and the bosonic fields is  modified to
\begin{equation}
i{\cal J}^{(1)}= 
     d\lambda^{(0)} +i\!\ast\! d\Lambda^{(D-2)}.\label{jnonconserv}
\end{equation}   

\bigskip

If the quantization of the fermion fields is compatible with the
conservation of the current ${\cal J}^{(1)}$, we will see that the scalar
field $\lambda$ is equal to zero and we recover the usual minimal
dualization.

Both (\ref{jconserv}) and (\ref{jnonconserv}) imply that for $D>2$,
the tensor field $\Lambda^{(D-2)}$ is not uniquely defined.  If two
$(D-2)$--forms $\Lambda$ and $\Lambda^\prime$ are related by
\begin{equation}
\Lambda^\prime - \Lambda= d\chi^{(D-3)},\label{gaugeform}
\end{equation}
for some antisymmetric $(D-3)$-form, $\chi^{(D-3)}$, then these
two $(D-2)$--forms yield the same current. The bosonic action has,
by construction, a gauge symmetry. The $(D-2)$--form $\Lambda$  is a
so--called ``gauge form" (generalization of a gauge field)~\cite{Frohlich}.

\bigskip

In Section 2, we will give a  full description of the extended
dualization method, and show that it permits the bosonization of
anomalous fermion theories in arbitrary dimension.

Extended dualization is a generalization of minimal dualization to
include anomalous quantum systems.  It also exhibits other virtues
that deserve mention here.  Within our extended approach, we can
reinterpret minimal dualization in a very simple way.  This
reinterpretation clearly shows that there is a considerable amount
of freedom involved in the dualization procedure.  This arbitrariness
in the bosonization is clearly differentiated from the arbitrariness
in the quantization of the fermion system.

Although minimal dualization  allows one to (at least in principle)
bosonize any arbitrary fermion theory with a dynamical conserved
current, it does not guarantee the corresponding bosonic action to
be easily tractable or even local.  The use of the above mentioned
arbitrariness in dualization, that  we identify in this paper,
can be very convenient in order to
find the most tractable form of bosonization.
For instance, we know that systems of non--relativistic fermions
at positive density yield a well behaved bosonic action using
minimal dualization~\cite{Frohlich}.  However, even for the relatively
simple case of a free massive Dirac fermion, the mass makes
dualization nontrivial.

\noindent The situation we will encounter in this paper is the following:
\begin{itemize} 
\item{}
For $D=2$, dualization  reproduces all the well known results of
conventional bosonization~\cite{Coleman,masspath,Joli}. The 
advantage of extended dualization is that it permits to bosonize
the most general quantization of the fermionic system.
\item{}     
For $D\ge 3$, we explore the low energy regime.  The resulting 
bosonic action turns out to be local. By contrast, it is not possible 
to obtain an equivalent bosonic formulation via dualization in the high 
energy regime. It is in this restricted sense that one has bosonization 
in arbitrary dimension.
\end{itemize}

\medskip
Dualization has the nice property that it permits one to obtain
relationships among a wide class of models.  We will apply this
property to obtain the bosonic equivalent of the $D$--dimensional
massive Thirring model (and other subsidiary models) very simply,
by adding some extra model--dependent bosonic terms to
the bosonization of the massive Dirac fermion.
We remark that, in contradistinction to minimal dualization,
extended dualization allows one to bosonize 
more general four--fermion interaction models including {\bf chiral} currents
(which are, in general, anomalous).

\noindent The dimension $D=2$ is a special case that
we consider separately in Section 2. As an example we discuss the
application of extended dualization to the bosonization of the 
Thirring model. We obtain the explicit dependence of the sine--Gordon 
equivalent interaction and the bosonic version of the currents on the 
parametrization of the quantization arbitrariness.

\bigskip
Section 3  will deal with the low energy regime, in
arbitrary dimension $D$. For large fermion mass, we  
study the bosonization of massive four--fermion interaction models.
The case of \underbar{chiral} four--fermion interactions is particularly
interesting because it can only be treated within the framework
of extended dualization.

One should mention here that some other work has been done in $D>2$
bosonization, for non--anomalous systems, 
using frameworks different from dualization.  In
particular, for the massive Thirring model in dimension $D=3$, 
the results in the abelian case~\cite{others} are a particular case  
of the bosonic action obtained in this paper using extended dualization.
On the other hand, the methods described in~\cite{others}, and their 
extension to the non--abelian case~\cite{Fradkinnon-ab}, are quite
convoluted. More importantly, those methods are very much tied to the
peculiarities of the Thirring model.  The dualization prescription,
by contrast, is a nice simple technique that allows one to
study a wide range of quantum fermion systems, among which the
Thirring model is just a particularly simple example.
Moreover, extended dualization is not restricted by any
symmetry requirement and it also allows the bosonization of 
anomalous systems. In section 3 we successfully complete the
bosonization of the \underbar{chiral} four--fermion interaction model.

\bigskip
Section 4 is devoted to the dualization of the simplest model in 2D
without non--anomalous quantization, the Chiral Schwinger model.
Extended dualization is very useful to show how quantization
ambiguities affect the bosonized version of the model.
An explicit parametrization of the quantization ambiguities associated
with the Chiral Schingwer model is exhibited.
In the bosonic version of the theory this quantization ambiguity
manifests itself as an additional scalar.
Moroever extended dualization permits one to write easily an 
explicit expression for the bosonic equivalent of the chiral current.

We will conclude, in Section 5, with a brief summary of the main
results on extended dualization.

\section{Extended dualization: Bo\-so\-ni\-za\-tion}

\subsection{ Extended dualization prescription}

Consider a quantum system of fermions in a flat euclidean 
spacetime\footnote{
In a general curved spacetime with non--trivial topology,
the extended dualization  procedure has to be modified,
as it is indicated in appendix A}
of arbitrary dimension $D$.  
The partition function is given by
\begin{equation} Z=\int {\cal D}\psi {\cal D}\overline\psi\quad
e^{-S_f(\psi,\overline\psi,\phi)},\label{fpart} \end{equation} where
$S_f$ includes any fermion self--interaction, and interactions
 with other external fields or source terms (generically denoted
by $\phi$).

We will start by describing a trick to modify the  partition function
(\ref{fpart}). It consists of the introduction of a path--integral
representation of the identity into the partition function.  We
call this trick ``extended dualization''.  Later we will establish
the relationship between such manipulations and the bosonization
of the fermion system.

\smallskip

Consider the following integral representation of the
identity:

\protect\begin{equation}
 1=\int {\cal D}A {\cal D}[\Lambda]{\cal D}\lambda\quad 
 e^{F(A,\psi,\overline\psi,\phi)}e^{(A,d\lambda+i\ast d\Lambda)}.
\label{identity}
\end{equation}
Here $A^{(1)}$ is an auxiliary $1$--form and the inner scalar product of 
forms\footnote{Given two rank $k$--forms $a^{(k)}={1\over
k!}a_{\mu_1...\mu_k}(x) dx^{\mu_1}\wedge ...\wedge dx^{\mu_k}$,
and $b^{(k)}$, we define the inner scalar product
\[
(a,b)\equiv \int d^D x\quad a_{\mu_1...\mu_k}(x) b^{\mu_1...\mu_k}(x) .
\]
Then 
\[
(A,d\lambda)=\int d^D x A^\mu(x)\partial_\mu\lambda(x),\quad
(A,i\!\ast\! d\Lambda)=
    i \epsilon_{\mu\mu_1...\mu_{D-1}}
    \int d^D x  A^\mu(x)
    \partial^{\mu_1}\Lambda^{\mu_2...\mu_{D-1}}(x).
\]} 
has been used.
$F(A,\psi,\overline\psi,\phi)$
is an arbitrary scalar functional depending on the original fields
and the non--physical vector field $A^{(1)}$, subject to the condition
\begin{equation}
F(A=0,\psi,\overline\psi,\phi)=0.\label{fcond}
\end{equation}
Furthermore, $\Lambda^{(D-2)}$ is an antisymmetric rank $(D-2)$ tensor
field and $\lambda^{(0)}$ is a rank $0$ scalar field.  ${\cal D}[\Lambda]$
denotes the measure on the space of gauge orbits
\begin{equation}
[\Lambda]=\lbrace \Lambda^\prime\colon
\Lambda^\prime - \Lambda= d\chi^{(D-3)}\rbrace.\label{orbits}
\end{equation}

\noindent We relegate the proof of (\ref{identity}) to Appendix A.
\bigskip

Now introduce the identity, in its path--integral form (\ref{identity}),
into the fermion partition function (\ref{fpart}). So far we have
nothing but an equivalence among path--integral expressions,
\begin{equation}
\int {\cal D}\psi {\cal D}\overline\psi\quad e^{-S_f}=
\int {\cal D}\psi {\cal D}\overline\psi
{\cal D}A {\cal D}[\Lambda]{\cal D}\lambda\quad 
e^{- S_f+F(A,\psi,\overline\psi,\phi)
+(A,d\lambda+i\ast d\Lambda)}.\label{nothing}
\end{equation}

\noindent Now we want to extract from this identity a bosonic system
in terms of which we could calculate Green functions of the fermionic
theory.  For the purpose of bosonization we will first suppose that
the change in the order of integration leaves results unaltered.
In the right hand side of (\ref{nothing}) we will try to perform
the integration over the fermions and the auxiliary field $A$, and
leave an expression in terms of $\Lambda$ and $\lambda$ 
(which will be the fields of the bosonized action).
 
\medskip

Define the bosonic action $S_b$,  
\begin{equation}
e^{-S_b(\Lambda,\lambda,\phi)}\equiv
\int{\cal D}A{\cal D}\psi {\cal D}\overline\psi\quad 
e^{- S_f+F(A,\psi,\overline\psi,\phi)+(A,d\lambda+i\ast d\Lambda)}.
\label{Sb}
\end{equation}
By construction, we have
\begin{equation}
Z=\int{\cal D}\psi {\cal D}\overline\psi
   \quad e^{-S_f(\psi,\overline\psi,\phi)}
 =\int  {\cal D}[\Lambda]{\cal D}\lambda\quad 
   e^{-S_b(\Lambda,\lambda,\phi)}.\label{Zequiv}
\end{equation}

\noindent Therefore, for any fermionic action 
$S_f(\psi,\overline\psi,\phi)$,
we have introduced an equivalent bosonic action 
$S_b(\Lambda,\lambda,\phi)$.  
The partition function of the quantum system, originally given in terms
of the fermionic field $\psi$, admits a bosonic representation in
terms of the fields $\Lambda^{(D-2)}$ and $\lambda^{(0)}$.

\noindent It immediately follows from its definition that the
bosonic action is invariant under an abelian gauge transformation of
the form $\Lambda\rightarrow\Lambda+d\chi^{(D-3)}$.  One says that the form
$\Lambda^{(D-2)}$ is a gauge form, which is to be integrated over the
space of gauge orbits.

\bigskip

So far, the bosonic action $S_b(\Lambda,\lambda,\phi)$ defined in (\ref{Sb})
includes an arbitrary functional $F(A,\psi,\overline\psi,\phi)$. It is
a matter of delicacy and judgement to choose $F$ in such a way that
the bosonic action, and the identification between Green functions
of the fermion and boson models, become tractable.

In the case of a fermion system whose interaction involves a single
current density ${\cal J}^{(1)}(\psi,\overline\psi)$, 
there is a particularly natural
and economic choice of the functional $F(A,\psi,\overline\psi,\phi)$:
\begin{equation}
F(A,\psi,\overline\psi,\phi)=-i(A,{\cal J}).\label{fchoice}
\end{equation}

\noindent The convenience of the choice (\ref{fchoice})  is clear
if, in (\ref{Sb}), one first integrates out the auxiliary field
$A$. The current then has a nice and simple bosonic equivalent
\begin{equation}
\int {\cal D}A\quad 
e^{-i (A,{\cal J})}e^{(A,d\lambda+i\ast d\Lambda)}
=\delta (-i{\cal J}+d\lambda+i\!\ast\! d\Lambda).\label{intA}
\end{equation}

\noindent Therefore, the bosonic action defined by
\begin{equation}
e^{-S_b(\Lambda,\lambda,\phi)}\equiv
\int{\cal D}A{\cal D}\psi {\cal D}\overline\psi\quad 
e^{- S_f-i(A,{\cal J})+(A,d\lambda+i\ast d\Lambda)},
\label{Sbj}
\end{equation}
makes the identification of the bosonic equivalent of the fermion
current density very simple
\begin{equation}
i{\cal J}(x)=
d\lambda(x)+i\!\ast\! d\Lambda(x).\label{jequiv}
\end{equation}

If the dynamics of the system involves several current densities 
${\cal J}^a$, the prescription (\ref{fchoice}) can be generalized by 
introducing as many auxiliary vector fields $A^a$, and bosonic fields 
$\lambda^a$ and $\Lambda^a$, as there are different dynamical currents. 
Then one has the identification (\ref{jequiv}) for each label $a$. For
example, this is one way to get a dual version of non--abelian systems
(such as fermions with non--abelian gauge interactions or non--abelian
Thirring models).

\bigskip
\noindent\underbar{Remark:}  Bosonization consists of establishing 
a relationship between fermionic and bosonic variables and, because 
of its generality, it cannot be a unique procedure. In this paper 
we explicitly demonstrate that behaviour. From the point of view of 
finding equivalent representations of the given partition function
(\ref{Zequiv}), other choices for the functional 
$F(A,\psi,\overline\psi,\phi)$
[different from (\ref{fchoice})] are equally valid. This arbitrariness
in dualization, which differs from the quantization arbitrariness
[inherent  in the fermionic integral in (\ref{Sb})], is related to
the freedom in choosing the bosonic variables (more precisely,
to the relationship between bosonic and fermionic variables). 

\noindent This function $F$ can be used to simplify
the bosonic action by, for example, cancelling bothersome terms in the
$A$ integration. This could be interesting when the fermionic
effective action is not quadratic in $A$ (as is the case 
for $D>2$ abelian systems beyond the low energy regime, and for 
$D$--dimensional non--abelian systems in any energy regime).
 
\noindent Even more, the function F can help us by providing a symmetry 
in the resulting bosonic action. For example, for $D=3$ the field
$\Lambda$ is an abelian gauge field. However, for non--abelian fermionic
systems this abelian gauge field can be converted into a non--abelian
one by playing with the arbitrariness in F~\cite{Schaposnik}.
We will go back to this point later on.

\noindent The price to pay for such extra freedom to choose $F$ is
that the identification of the bosonic equivalent of the current
density will be more complicated than (\ref{jequiv}). This means that
we are choosing other bosonic variables not so trivially related to 
the fermionic variables, but better adapted to the bosonic version
of the theory.

\noindent In this paper we will illustrate extended dualization in
simple cases where it is not necessary to use a nontrivial choice
of bosonic variables.

\subsection{Properties of bosonization by extended dualization}
 
Let us concentrate on the properties of abelian bosonization defined
by relation (\ref{Sbj}).

\newtheorem{proposition}{Property}[subsection]
 
\begin{proposition}
$S_f$ and $S_b$ related by (\ref{Sbj}), describe the same partition
function,
\begin{equation}
Z=\int {\cal D}\psi {\cal D}\overline\psi\quad 
e^{-S_f(\psi,\overline\psi,\phi)}=
\int {\cal D}[\Lambda]{\cal D}\lambda\quad 
e^{- S_b(\Lambda,\lambda,\phi)}.\label{partfunct}
\end{equation}
\end{proposition}
This property follows by construction (assuming the order of
integration can be exchanged).

\bigskip

The expression for the fermi fields in the bosonic theory will, in
general, be very complicated~\cite{Mandelstam,Frohlich}. However,
fermion bilinears like ${\cal J}^{(1)}$ have simple expressions.

\begin{proposition}
Any correlation function of the current $i{\cal J}(x)$ in the fermionic
theory is related to a correlation function of $\lbrack
d\lambda(x)+i\!\ast\! d\Lambda(x)\rbrack$ in the bosonic model by
\begin{equation}
\langle i{\cal J}^{\mu_1}(x_1)...i{\cal J}^{\mu_n}(x_n)\rangle_f=
\langle {\lbrack d\lambda+i\!\ast\! d\Lambda}\rbrack^{\mu_1}(x_1)...
{\lbrack d\lambda+i\!\ast\! d\Lambda}\rbrack^{\mu_n}(x_n)\rangle_b.
\label{currcorr}
\end{equation}
\end{proposition}

\noindent The proof is very straightforward 
by doing some path--integral manipulations
and changing the order of integrals.

\begin{proposition}
Consider a fermion--boson dual description of a given quantum
system, where the  actions  $S_f(\psi,\overline\psi,\phi)$ and 
$S_b(\Lambda,\lambda,\phi)$ are related by extended dualization.
 Any modified system obtained 
 from the original one by  adding
  current interaction terms in the fermionic  description
$S_f^\prime=S_f+S_{int}(i{\cal J})$
has a bosonic dual given by 
$S_b^\prime=S_b+S_{int}(d\lambda+i\ast d\Lambda)$
\end{proposition}

\noindent As a consequence of this property we can trivially 
dualize a large class of systems obtained from a given one, by 
adding suitable current density interaction terms.

\noindent Examples are,
\begin{itemize}
\item{}
Current--current interaction terms, such as
\begin{equation}
  S_{int}=+{1\over 2}\int d^D x d^D y {\cal J}_\mu(x)
  {V}_{\mu\nu}(x,y){\cal J}_\nu(y)
  =+{1\over 2}({\cal J},V{\cal J}),\label{currentcurrent}
\end{equation}
modify the original bosonized action $S_b$ by the term
\begin{equation}
  -{1\over 2}\left({ d\lambda+i\!\ast\! d\Lambda},
  {V}
  {\left[ d\lambda+i\!\ast\! d\Lambda\right]}\right).\label{dualcurrent}
\end{equation}
Therefore, quartic interactions of fermions become quadratic in 
terms of the bosonic fields.
\item{} 
Source terms for the currents, $j^{(1)}$, and interactions with
abelian gauge fields $a^{(1)}$, of the form
\begin{equation}
  S_{int}=(i{\cal J},j+a),\label{source}
\end{equation}
\noindent lead in the bosonic action to
\begin{equation}
  (d\lambda+i\!\ast\! d\Lambda,j+a).\label{dualsource}
\end{equation}
\end{itemize}

\begin{proposition}
If the quantum field theory admits a quantized  dynamical  conserved
fermion current density, then the scalar field $\lambda^{(0)}$
is equal to zero, and one recovers minimal dualization.
\end{proposition}

\noindent\underline{Proof:}

\noindent Let ${\cal J}^{(1)}(\psi,\overline\psi)$ 
be a quantized dynamical conserved 
current density of the system, and let us use this current density to 
define the bosonized version of the theory:
\begin{equation}
e^{-S_b(\Lambda,\lambda)}=
\int {\cal D}A{\cal D}\psi{\cal D}\overline\psi\quad
e^{-S_f -i (A,{\cal J})+(A,d\lambda+i\ast d\Lambda)}. 
\label{Sgamma}
\end{equation}
\noindent Now define the ``effective action"
\begin{equation}
e^{-\Gamma(A)}\equiv\int {\cal D}\psi{\cal D}\overline\psi\quad
e^{-S_f -i (A,{\cal J})}. \label{Gamma}
\end{equation}

\noindent First note that, with this dualization, the quantum
current density is conserved iff the effective action is gauge
invariant.
\begin{equation}
d\!\ast\!{\cal J}^{(1)}=0\quad\Longleftrightarrow 
\Gamma(A)=\Gamma(A+d\alpha).\label{gaugeinv}
\end{equation}

\noindent Let us study the consequences of this gauge symmetry on
the bosonic action,
\begin{equation}
 e^{-S_b(\Lambda,\lambda)}=\int {\cal D}A\quad
 e^{-\Gamma(A)+(A,d\lambda+i\ast d\Lambda)}.
\label{Sinv}
\end{equation}
Any non--harmonic one-form $A^{(1)}$ can be cast into the form
\begin{equation}
A^{(1)}= d\zeta^{(0)}  + i\!\ast\! d \varphi^{(D-2)}.\label{descomp}
\end{equation}
In terms of the fields $\zeta^{(0)}$ and $\varphi^{(D-2)}$ we have,
\begin{eqnarray}
 e^{-S_b(\Lambda,\lambda)}
 &\propto&\int {\cal D}\zeta{\cal D}\varphi\quad e^{-\Gamma(\varphi)
 +(d\zeta+i \ast d\varphi,d\lambda+i\ast d\Lambda)} 
\label{Seta}\\
 &=& \int {\cal D}\zeta\quad e^{(d\zeta,d\lambda)}
 \int {\cal D}\varphi\quad e^{-\Gamma(\varphi)
  -(d\varphi,d\Lambda)} 
\nonumber\\
 &\propto& \delta(\lambda)\int {\cal D}\varphi\quad 
 e^{-\Gamma(\varphi)-(d\varphi,d\Lambda)}  
\nonumber\\
 &\propto& \delta(\lambda)\int {\cal D}[A]\quad e^{-\Gamma(A)
 + (A,i\ast d\Lambda)}. 
\nonumber
\end{eqnarray}
Here ${\cal D}[A]$ is the integral over the space of gauge orbits.

\noindent The scalar field $\lambda^{(0)}$ trivially disappears because
it is set to zero by the delta function.  We can use
\begin{equation}
e^{-S_b(\Lambda)}\equiv\int {\cal D}[A]\quad e^{-\Gamma(A)
 + (A,i\ast d\Lambda)}, 
\end{equation}
to describe the partition function, originally introduced in terms
of fermions.  This is the result of minimal
dualization~\cite{BuQue2d,Frohlich,BuQue}.

 \bigskip

\noindent\underbar{Remark:} Note that the extended dualization
presented in this paper, first enlarges minimal dualization because
it permits the discussion of anomalous theories. Second, simplifies
minimal dualization because it is introduced by an integral
representation of the identity. And third, it allows to identify
a big freedom in the dualization procedure (associated to the choice
of the functional $F$). This arbitrariness is clearly differentiated
from the possible arbitrariness in the path--integral quantization.

Another advantage of extended dualization is that, in contrast with
minimal dualization, one does not need to use any symmetry of the 
fermion system, even if such symmetry does exist. This fact opens 
new possibilities for bosonizing fermionic theories.

\noindent Perhaps the most remarkable examples where we can exploit 
this advantage are non--abelian systems. The natural extension of 
minimal dualization to this case involves  non--abelian gauge fields 
$A^a$ and tensor fields $\Lambda^a$. This last field acts as a Lagrange 
multiplier for the condition ${\cal F}^a_{\mu\nu} (A) = 0$ and makes
the gauge field trivial due to the non--abelian symmetry of the 
system~\cite{BuQue2dnon-ab,Schaposnik}. The problem is now, that, the 
coupling between $A$ and $\Lambda$ is not linear in $A$ and therefore
it is not possible an exact identification of the fermionic current
like (\ref{jconservcomp}). Due to this fact one loses all the usual
properties such as the direct relation between Green functions 
(property 2.2.2) or the simple rule to introduce interactions
(property 2.2.3).

 However, extended dualization of non--abelian systems does
not require any modification because one does not need to implement
any non--abelian
symmetry in the theory. As we said previously, one 
can apply our dualization prescription (\ref{Sbj}) independently to
each current density ${\cal J}^{a}$ with the choice (\ref{fchoice})
for $F$ which is linear in $A^a$. Proceeding in this way, all the 
properties of bosonization by extended dualization remain in the 
non--abelian case. We should remark here that this prescription
would lead to an independent abelian gauge
symmetry for each field $A^a$, in contradistinction with the
non--abelian symmetry of minimal dualization.

The above prescription leads to a bosonic action with an abelian
gauge symmetry for each field $\Lambda^a$. For $D=3$ $\Lambda$ is a
vector field; if we want for the bosonic action to have a non--abelian
symmetry with $\Lambda^a_\mu$ as a gauge field, extended dualization 
is again needed. The coupling
between $\Lambda^a_\mu$ and ${\cal F}^a_{\mu\nu} (A)$ is not
gauge invariant and one needs to use the freedom in the function $F$
to get such non--abelian symmetry. 
It has been shown~\cite{Schaposnik} that the mixed Chern--Simons
term
\begin{equation}
  {i\over 2\pi}\int d^3 x \quad\epsilon^{\mu\nu\rho}\Lambda^a_\mu
			 {\cal F}^a_{\nu\rho}(A)
-{i\over 24\pi}\int d^3 x \quad\epsilon^{\mu\nu\rho}f_{abc}
			 A^a_\mu A^b_\nu A^c_\rho,
\end{equation}                     
changes by an integer multiple of $2\pi i$ under simultaneous
non--abelian gauge transformations of $\Lambda^a_\mu$ and
$A^a_\mu$. That is, a choice of $F(A)$ including a term proportional
to 
\begin{equation}
i\int d^3 x \quad\epsilon^{\mu\nu\rho}f_{abc} A^a_\mu A^b_\nu A^c_\rho,
\end{equation}
cancells the gauge non--invariance of the coupling between
 $\Lambda^a_\mu$ and  ${\cal F}^a_{\mu\nu} (A)$ (up to an irrelevant
 multiple of $2\pi i$).
 In this way, one gets a bosonic action with a non--abelian
 symmetry as we had in the fermionic theory. Of course, with
 this prescription, we pay the price of loosing the above mentioned
 properties of bosonization.

\noindent For simplicity only abelian examples will be treated in 
this paper. A detailed analysis of non--abelian systems will be the
subject of another paper.

\subsection{$D=2$ free massless Dirac fermion and related mo\-dels}

The conventional bosonization of the $D=2$ free Dirac fermion is
well known~\cite{Abdalla}.  It  is obtained by using a quantization 
where the vector--like fermion current density 
$\overline\psi\gamma_\mu\psi$ 
is conserved.  The result is that the free Dirac fermion in $D=2$ is
equivalent to a free scalar boson $\Lambda$.  The bosonized version of
the vector--like fermion current is 
$\epsilon_{\mu\nu}\partial^\nu\Lambda$.
Minimal dualization agrees with conventional
bosonization~\cite{BuQue2d,Frohlich}. However, using all the
arbitrariness in the quantization, one gets a gauge non--invariant
result for the fermionic path--integral~\cite{Fuji,Das}. Using extended 
dualization it is possible to obtain an explicit parametrization of the 
quantization arbitrariness in the bosonized version of the system.

The partition function is defined by
\begin{equation}
Z_{free}=\int {\cal D}\psi{\cal D}\overline\psi\quad
e^{-S_f^{free}}.\label{freeZ}
\end{equation}
Here $S^{free}_f(\psi,\overline\psi)$ stands for the free Dirac action
in flat euclidean spacetime\footnote{Our conventions are:
\[
\gamma_\mu^\dagger=\gamma_\mu,\quad
\lbrace\gamma_\mu,\gamma_\nu\rbrace=2\delta_{\mu\nu},\quad
\gamma_5^\dagger= \gamma_5.
\]
In $D=2$ we use \(\gamma_\mu\gamma_5=i\epsilon_{\mu\nu}\gamma^\nu.\)}
\begin{equation}
 S^{free}_f(\psi,\overline\psi)=
 -\overline\psi\gamma^\mu\partial_\mu\psi.\label{sfree}
\end{equation}

In $D=2$ the antisymmetric form $\Lambda^{(D-2)}$ reduces to a scalar
field (a zero-form).  The bosonic action obtained by dualization
is given in terms of the two scalar fields $\Lambda$ and $\lambda$ by
\begin{eqnarray}
e^{-S^{free}_{b}(\Lambda,\lambda)}
 &=&\int {\cal D}A{\cal D}\psi{\cal D}\overline\psi\quad
 e^{-S^{free}_f-i (A,{\cal J})+(A,d\lambda+i\ast d\Lambda)}
\label{freeb}\\
 &=&\int {\cal D}A{\cal D}\psi{\cal D}\overline\psi\quad
 e^{\overline\psi\gamma^\mu(\partial_\mu-i A_\mu)\psi
 +(A,d\lambda+i\ast d\Lambda)}.\nonumber
\end{eqnarray}
Here we have dualized using the vector-like current ${\cal J}_\mu(x)=
\overline\psi(x)\gamma_\mu\psi(x)$.
 
We introduce the effective action,
\begin{equation}
 e^{-\Gamma(A)}=\int{\cal D}\psi{\cal D}\overline\psi\quad 
 e^{\overline\psi\gamma^\mu(\partial_\mu-i A_\mu)\psi},\label{freegamma} 
\end{equation}
in terms of which the bosonic action reads
\begin{equation}
 S^{free}_{b}(\Lambda,\lambda)=-\log\int{\cal D}A\quad 
 e^{-\Gamma(A)+(A,d\lambda+i\ast d\Lambda)}.\label{gammasbfree} 
\end{equation}

The effective action $\Gamma(A)$ can be calculated exactly. However,
we will use a symmetry argument to obtain its value.  Consider a
local and arbitrary transformation over the fermion fields
\begin{eqnarray}
\psi&\longrightarrow& 
e^{i(\alpha+\beta\gamma_5)}\psi,\label{transf}\\ 
\overline\psi&\longrightarrow& 
\overline\psi e^{-i(\alpha-\beta\gamma_5)}.\nonumber
\end{eqnarray}

\noindent Under such transformation the effective action has the
following property
\begin{equation}
\Gamma(A)=\Gamma(A-d\alpha+i \ast d\beta)-\log J(\alpha,\beta,A),
\label{gammaprop}
\end{equation}
where $J(\alpha,\beta,A)$ is the finite Jacobian~\cite{Fuji} 
corresponding to the fermion transformation (\ref{transf}).

\noindent The most general expression for this Jacobian can be
read, for example, from~\cite{Das}.  It has the form
\begin{equation}
\log J(\alpha,\beta,A)={i\over\pi}\left[
\xi(\beta,\epsilon_{\mu\nu}\partial_\mu A_\nu)
+i\eta(\alpha,\partial_\mu A_\mu)\right]
-{1\over 2\pi}\left[  \xi(d\beta,d\beta)
+\eta(d\alpha,d\alpha)\right].\label{dasj}
\end{equation}
Here $\xi$ and $\eta$ are parameters introduced to describe the
particular quantization employed.  They satisfy the relation
$\xi+\eta=1$.

\noindent The particular case ($\eta=0$, $\xi=1$) corresponds to
a gauge invariant quantization. This special case is equivalent to
minimal dualization~\cite{BuQue2d,Frohlich}.

Property (\ref{gammaprop}) implies that the bosonic action satisfies
\begin{eqnarray}
S^{free}_{b}\left(\Lambda,\lambda\right)&=& 
S^{free}_{b}\left(\Lambda+{\xi\over\pi}\beta,
       \lambda+{\eta\over\pi}\alpha\right)\label{sbprop}\\
&&-{1\over 2\pi}\left[ \xi(d\beta,d\beta)
+\eta(d\alpha,d\alpha)\right]
-(d\beta,d\Lambda)-(d\alpha,d\lambda).\nonumber
\end{eqnarray}

\noindent As a consequence: 
\begin{eqnarray}
S^{free}_{b}(\Lambda,\lambda)-{\pi\over 2\xi}(d\Lambda,d\Lambda)
& 
\mbox{\rm is invariant under}
&
\Lambda\rightarrow \Lambda+{\xi\over\pi}\beta,
\quad\lambda\rightarrow\lambda,\label{sbsim}
\nonumber\\
S^{free}_{b}(\Lambda,\lambda)-{\pi\over 2\eta}(d\lambda,d\lambda)
& 
\mbox{\rm is invariant under}
&
\lambda\rightarrow \lambda+{\eta\over\pi}\alpha,
\quad\Lambda\rightarrow\Lambda.
\nonumber\\
&&
\end{eqnarray}

\noindent Therefore we get the final result
\begin{equation}
  S^{free}_{b}(\Lambda,\lambda)={\pi\over 2\xi}(d\Lambda,d\Lambda)
 +{\pi\over 2\eta}(d\lambda,d\lambda)\label{sfinal}
\end{equation}

Scaling to canonical variables, 
$\Lambda\rightarrow\sqrt{\xi\over\pi}\Lambda$
and $\lambda\rightarrow\sqrt{\eta\over\pi}\lambda$, 
we find that the free
Dirac fermion in $D=2$ bosonizes into two free scalars. By construction
we have the following identification for the fermion currents
\begin{eqnarray}
 \overline\psi\gamma_\mu\psi&\longleftrightarrow&
 \sqrt{{\xi\over\pi}}\epsilon_{\mu\nu}\partial^\nu\Lambda
 -\sqrt{\eta\over\pi}i\partial_\mu\lambda,\label{j2dfree}\\
 \overline\psi\gamma_\mu\gamma_5\psi&\longleftrightarrow&
 \sqrt{\eta\over\pi}\epsilon_{\mu\nu}\partial^\nu\lambda
 - \sqrt{{\xi\over\pi}}i\partial_\mu\Lambda.\nonumber
\end{eqnarray}

\noindent The gauge invariant result corresponds to $\lambda=0$ 
and $\xi=1$.

Other related models can be trivially bosonized once we have
expression (\ref{sfinal}) in hand.  In particular, consider 
the fermionic action given by
\begin{equation}
  S_f=S^{free}_f
  +(i{\cal J},j)+{g\over 2}({\cal J},{\cal J}),\label{sextend}
\end{equation}
with a source term $j^{(1)}$ for the current density, and a
Thirring--like interaction  of strength $g$.

\noindent Using the general properties of extended dualization 
we have
\begin{eqnarray}
  S_b(\Lambda,\lambda)\!\!&\!=\!\!&\!S^{free}_b(\Lambda,\lambda)
  +(j,d\lambda+i\ast d\Lambda)
  -{g\over 2}(d\lambda+ 
     i\!\ast\!d\Lambda,d\lambda+i\!\ast\!d\Lambda)\label{sbextend} \\
  \!&=&\!\!{1\over 2}\left({\pi\over\xi}+g\right)(d\Lambda,d\Lambda)
  +{1\over 2}\left({\pi\over\eta}-g\right)(d\lambda,d\lambda)
  +(j,d\lambda+i\!\ast\!d\Lambda).\nonumber
\end{eqnarray}
 
\medskip

\noindent The net effect of the fermion self--interaction is a 
change of the kinetic coefficients for the scalar fields.
  
\noindent The stability of the Thirring model requires
\begin{equation}
  -{\pi\over\xi}<g<{\pi\over\eta}.\label{thstability}
\end{equation}
This is a generalization of the usual condition 
$-\pi<g$. See~\cite{Coleman}.

\noindent In the range of stability, we rescale $\Lambda$ and $\lambda$ to
canonical variables
\begin{equation}
   \Lambda\rightarrow {\beta\over 2\pi}\Lambda,\quad\quad 
  \lambda\rightarrow  {\gamma\over 2\pi}\lambda,\label{lchange}
\end{equation}
 where $\beta$ and $\gamma$ are constants defined by
\begin{equation}
   {4\pi\over \beta^2} = {1\over\xi}+{g\over\pi},\quad\quad
  {4\pi\over \gamma^2} = {1\over\eta}-{g\over\pi}.\label{bgdef}
\end{equation}
   
\medskip

\noindent In terms of the canonical variables we have
\begin{equation}     
 S_b(\Lambda,\lambda)={1\over 2}(d\Lambda,d\Lambda)
  +{1\over 2}(d\lambda,d\lambda)
  +(j,{\gamma\over 2\pi}d\lambda+
  {\beta\over 2\pi}i\!\ast\! d\Lambda).\label{sbcanonical}
\end{equation}
  
\noindent The identification of currents is
\begin{eqnarray}
 \overline\psi\gamma_\mu\psi&\longleftrightarrow&
 {\beta\over 2\pi}\epsilon_{\mu\nu}\partial^\nu\Lambda
 -{\gamma\over 2\pi}i\partial_\mu\lambda,\label{current2dth}\\
 \overline\psi\gamma_\mu\gamma_5\psi&\longleftrightarrow&
 {\gamma\over 2\pi}\epsilon_{\mu\nu}\partial^\nu\lambda
 -{\beta\over 2\pi}i\partial_\mu\Lambda.\nonumber
\end{eqnarray}

The particular case ($\beta^2={4\pi\xi}$, $\gamma^2={4\pi\eta}$)
corresponds to the free Dirac fermion.

\subsection{$D=2$ free massive Dirac fermion and re\-la\-ted mo\-dels}

The partition function is defined by
\begin{equation}
Z_{free}(m)=\int {\cal D}\psi{\cal D}\overline\psi\quad
e^{-S^{free}_f(m;\psi,\overline\psi)},\label{freemZ}
\end{equation}
while the free massive Dirac action has the form
\begin{equation}
 S^{free}_f(m;\psi,\overline\psi)=-
 \lbrack\overline\psi\gamma^\mu\partial_\mu\psi
 +m\overline\psi\psi\rbrack. \label{sfreem}
\end{equation}

Consequently, the bosonic action for a free massive Dirac fermion
has the following path integral expression
\begin{eqnarray}
S^{free}_{b}(m;\Lambda,\lambda)
 &=&\int{\cal D}A{\cal D}\psi{\cal D}\overline\psi\quad
 e^{-S^{free}_f-i (A,{\cal J})+(A,d\lambda+i\ast d\Lambda)}
\label{freebm}\\
 &=&\int{\cal D}A{\cal D}\psi{\cal D}\overline\psi\quad
 e^{\overline\psi\gamma^\mu(\partial_\mu-i A_\mu)\psi
 +m\overline\psi\psi 
 +(A,d\lambda+i\ast d\Lambda)}.\nonumber
\end{eqnarray}
 
Now, introduce the effective action,
\begin{equation}
 e^{-\Gamma_m(A)}=\int{\cal D}\psi{\cal D}\overline\psi\quad 
 e^{\overline\psi\gamma^\mu(\partial_\mu-i A_\mu)\psi
 +m\overline\psi\psi},\label{freegammam} 
\end{equation}
in terms of which
 
\begin{equation}
 S^{free}_{b}(m;\Lambda,\lambda)=\int{\cal D}A\quad 
 e^{-\Gamma_m(A)+(A,d\lambda+i\ast d\Lambda)}.\label{gammasbfreem} 
\end{equation}

The bosonic action $S^{free}_b(m;\Lambda,\lambda)$ can
be calculated by a power expansion in the fermionic mass.  This
procedure is well known in the $D=2$ bosonization
folklore~\cite{masspath,Joli}.  The main steps of the calculation
are exhibited in Appendix B, and the final result is
\begin{eqnarray}
S^{free}_b(m;\Lambda,\lambda)&=&S^{free}_b(0;\Lambda,\lambda)
-2m\Lambda_{UV}\int d^2 x\cos\left[ {2\pi\over\xi}\Lambda(x)\right]
\label{2dmsbfree}\\
  &=&{\pi\over 2\xi}(d\Lambda,d\Lambda)
 +{\pi\over 2\eta}(d\lambda,d\lambda)
 -2m\Lambda_{UV}\int d^2 x\cos\left[ {2\pi\over\xi}\Lambda(x)\right].
 \nonumber
\end{eqnarray}
Here $\Lambda_{UV}$  is an ultraviolet cutoff required to regularize
the integral over the  auxiliary  field $A$.
 
\medskip

Adding other interactions and sources to the bosonized free
massive Dirac fermion is, as before, trivial. Take the fermionic
action to be
\begin{equation}
S_f=S^{free}_f(m;\overline\psi,\psi)
+(i{\cal J},j)+{g\over 2}({\cal J},{\cal J}).\label{smextend}
\end{equation}
The corresponding bosonic action is given by
\begin{eqnarray}
S_b(m;\Lambda,\lambda,j)&=&
  {1\over 2}\left({\pi\over\xi}+g\right)(d\Lambda,d\Lambda)
  +{1\over 2}
     \left({\pi\over\eta}-g\right)(d\lambda,d\lambda)\label{sbmfree}\\
 &&-2m\Lambda_{UV}\int d^2 x\cos\left[ 
 {2\pi\over\xi}\Lambda(x)\right]
  +(j,d\lambda+i\!\ast\!d\Lambda).\nonumber
\end{eqnarray}
Rescaling to the canonical variables introduced in (\ref{lchange}),
(\ref{bgdef}), we have
 \begin{eqnarray}     
 S_b(m;\Lambda,\lambda,j)&=&{1\over 2}(d\Lambda,d\Lambda)
  +{1\over 2}(d\lambda,d\lambda) \label{sbm2dth}\\ 
 &&-2m\Lambda_{UV}\int d^2 x\cos\left[ 
 {\beta\over\xi}\Lambda(x)\right]
  +(j,{\gamma\over 2\pi}d\lambda+
  {\beta\over 2\pi}i\!\ast\! d\Lambda).\nonumber
\end{eqnarray}
The identification of the fermion current densities is not modified
from the massless case (\ref{current2dth}), neither is the stability
condition (\ref{thstability}).

This is a generalization of the well known result~\cite{Coleman,masspath}
that the massive Thirring model is equivalent
to the  sine--Gordon model.  In our analysis there is also an
additional scalar field $\lambda$, that plays a role in the identification
of the fermionic current densities
\begin{eqnarray}
 \overline\psi\gamma_\mu\psi&\longleftrightarrow&
 {\beta\over 2\pi}\epsilon_{\mu\nu}\partial^\nu\Lambda
 -{\gamma\over 2\pi}i\partial_\mu\lambda,\label{current2dthm}\\
 \overline\psi\gamma_\mu\gamma_5\psi&\longleftrightarrow&
 {\gamma\over 2\pi}\epsilon_{\mu\nu}\partial^\nu\lambda
 -{\beta\over 2\pi}i\partial_\mu\Lambda.\nonumber
\end{eqnarray}

\noindent The parameter values $\beta^2=4\pi\xi$, and $\gamma^2=4\pi\eta$,
correspond to the massive free Dirac fermion system. In particular,
setting $\beta^2=4\pi$, and $\gamma^2=0$, corresponds to minimal
dualization.

\section{$D$--dimensional massive four--fermion interactions}
In this section we will concentrate on the dualization of D--dimensional
massive four--fermion interactions. First of all, we shall formally
derive the bosonic action by invoking the bosonization of the free massive
Dirac fermion and then incorporating extra bosonic terms. These extra 
terms will depend both on the model and on the particular dualization
prescription.
Later on, we shall review the results obtained for the massive Thirring
model using minimal dualization in the low energy limit. Finally,
we will explore the potential advantages of extended dualization by
performing the bosonization of the chiral four--fermion interaction
model (low energy only).

For definiteness
\begin{equation}
 S_f^{4f}(m;\psi,\overline\psi)=S^{free}_f(m;\psi,\overline\psi)
 +{g\over 2} ({\cal J},{\cal J}).
 \label{sth}
\end{equation}
Here $S^{free}_f(m;\psi,\overline\psi)$ is the free massive Dirac action
defined in (\ref{sfreem}), while the density current ${\cal J}^{(1)}(x)$ 
is a rank--one fermion bilinear.

The quantum system is defined by the partition function.
Including  a source $j^{(1)}$ for the current, this reads
\begin{equation}
Z_{4f}(j)=\int {\cal D}\psi{\cal D}\overline\psi\quad
e^{-S_f^{4f}- (i{\cal J},j)}.\label{thZ}
\end{equation}

The extended dualization of this model proceeds as follows
\begin{equation}
Z_{4f}(j)=\int{\cal D}[\Lambda]{\cal D}\lambda
     \quad e^{-S_b^{4f}(m;\Lambda,\lambda,j)}.\label{thdual}
\end{equation}

\noindent The bosonic action $S_b^{4f}$ is given  by
\begin{equation}
S_b^{4f}(m;\Lambda,\lambda,j)= S^{free}_{b}(m;\Lambda,\lambda)
+{g\over 2}\left[(d\Lambda,d\Lambda)-(d\lambda,d\lambda)\right]
+(j,d\lambda+i\ast d\Lambda).\label{sbth}
\end{equation}

\noindent Therefore, to obtain the bosonic representation of the
massive four--fermion interaction model, 
the only calculation left to do is the evaluation of
bosonic action for the free massive Dirac fermion
$S^{free}_{b}(m;\Lambda,\lambda)$
\begin{equation}
S^{free}_{b}(m;\Lambda,\lambda) =\int{\cal D}A\quad 
 e^{-\Gamma_m(A)+(A,d\lambda+i\ast d\Lambda)}.\label{sbgammam} 
\end{equation}

In general, for arbitrary mass and dimension,
the effective action $\Gamma_m(A)$, given by (\ref{freegammam}),  
is a very complicated functional. However, in the
limit of large fermionic mass, $m\rightarrow\infty$, the effective
action becomes quadratic in the field $A$ and we can give an explicit
expression for the bosonic action 
$S^{free}_{b}(m\nearrow\infty;\Lambda,\lambda)$.

\subsection{Massive Thirring model: Low energy limit}

The  massive Thirring model is one of the simplest four--fermi
interaction systems suitable for dualization. Other approaches have
been used to bosonize the massive Thirring model~\cite{others}.
However, dualization makes the bosonization  a lot simpler.
Additionally, while other approaches rely on rather specific
properties of the Thirring model,  dualization applies to any fermionic
system.

In the Thirring model the current density, 
${\cal J}_\mu(x)= \overline\psi\gamma_\mu\psi$ is vector--like. 
Therefore, the system admits a quantization
that conserves the vector--like current and we can apply the minimal
dualization prescription.

In the large mass limit, the gauge invariant effective action
$\Gamma_m(A)$ is quadratic in the field $A$, and can be cast into
the form
\begin{eqnarray}
\Gamma_m(A)&=&{1\over 2}\int d^D x\quad
A^\mu(x)C^D_{\mu\nu}(\partial,m)A^\nu(x)\label{invlargem}\\
&=&{1\over 2}(A,C^D A).\nonumber
\end{eqnarray}
Using the well known results for the differential operator 
$C^D_{\mu\nu}(\partial,m)$~\cite{BuQue} ,
we can perform the gaussian integral
in (\ref{sbgammam}) in an appropiate gauge.
\bigskip

\noindent The final results for the  Thirring model in the 
infinite mass limit are:
\begin{description}
\item[D=2] 
The bosonic action has a scalar field $\Lambda$ with a mass that is
proportional to the fermion mass and to the inverse of the 
Thirring coupling $g$.  It is given by
\begin{equation}
S_b^{(Th)}(m\nearrow\!\infty;\Lambda,j)= {1\over 2}
\left[g+{6\pi \over 5}\right](d\Lambda,d\Lambda)
+{1\over 2}6\pi m^2(\Lambda,\Lambda)+i(j,\ast d\Lambda).\label{2th}
\end{equation}
The mass for the scalar $\Lambda$ is $M^2=5 m^2/ [1+{5g\over 6\pi}]$.

\item[D=3] 
The bosonic action includes a gauge field $\Lambda_\mu$, with an 
abelian Chern--Simons term
\begin{equation}
S_b^{(Th)}(m\nearrow\!\infty;\Lambda,j)= {1\over 2}g(d\Lambda,d\Lambda)
+{8\pi^2\over sign(m)}(\Lambda,\ast d\Lambda)
+i(j,\ast d\Lambda).\label{3th}
\end{equation}
The gauge field $\Lambda_\mu$ has a topological mass~\cite{DJT} given
by $M=8\pi^2/g$.

\item[D$\ge$ 4] 
The bosonic field is a rank $(D-2)$ antisymmetric form
$\Lambda_{\mu_1...\mu_{D-2}}$, and the action becomes
\begin{equation}
S_b^{(Th)}(m\nearrow\!\infty;\Lambda,j)= {1\over 2}g(d\Lambda,d\Lambda)
+{1\over  K_D}(\Lambda,\Lambda)+i(j,\ast d\Lambda).\label{4th}
\end{equation}
Therefore, in terms of the gauge form $\Lambda^{(D-2)}$, the bosonic
action is local in the infinite mass limit. The mass for the form 
$\Lambda$ is $M^2= 1/ g K_D$, where $K_D$ is a coefficient depending on
the regularization method.

\end{description}

The interpretation of these results as an effective theory at low
energies (much smaller than the fermion mass) will be discussed
elsewhere~\cite{Jorge}.

\subsection{Chiral four--fermion interactions: Low energy limit}
A considerably less trivial system is obtained when the interaction
is not parity invariant.
For definiteness specialize the current in (\ref{sth}) to be
${\cal J}_\mu= \overline\psi\gamma_\mu{1-\gamma_5\over 2}\psi$.
In this case, the relevant current of the interaction cannot be
quantized in a conserved way. We therefore need to invoke extended
dualization and thereby demonstrate its utility by successfully
completing the bosonization of this chiral model.

The most general result for the effective action
in the infinite mass limit is
\begin{eqnarray}
\Gamma_m(A)&=&
   -\log\int {\cal D}\psi{\cal D}\overline\psi\quad 
   e^{\overline\psi\gamma^\mu(\partial_\mu-
   i{1-\gamma_5\over 2} A_\mu)\psi
   +m\overline\psi\psi},\nonumber\\
&=&{1\over 2}{\kappa_D}(A,A),\label{largem}
\end{eqnarray}
with a regularization dependent and dimension dependent 
coefficient ${\kappa_D}$.

The bosonic action is given by
\begin{eqnarray}
S_b^{(chiral)}(m\nearrow\!\infty;\Lambda,\lambda,j)&=& 
{g\over 2}\left[(d\Lambda,d\Lambda)-(d\lambda,d\lambda)\right]
+i(j,d\lambda+i\ast d\Lambda)\nonumber\\
&&-\log\int{\cal D}A\quad e^{-{1\over 2}{\kappa_D}(A,A)
+(A,d\lambda+i\ast d\Lambda)},\label{ssthA}
\end{eqnarray}
where $j^{(1)}$ is a source for the chiral current.

\noindent The bosonic action contains two fields $\Lambda^{(D-2)}$ 
and $\lambda^{(0)}$ 
\begin{equation}
S_b^{(chiral)}(m\nearrow\!\infty;\Lambda,\lambda,j)= 
{1\over 2}\left(g+{1\over \kappa_D}\right)
\left[(d\Lambda,d\Lambda)-(d\lambda,d\lambda)\right]
+(j,d\lambda+i\ast d\Lambda).\label{ssthAs}
\end{equation}

This bosonic action is local and linear in the source, so we can
establish the operator identification
\begin{eqnarray}
i\overline\psi\gamma_\mu {1-\gamma_5\over 2}\psi
&\longleftrightarrow& 
d\lambda+i\ast d\Lambda. \label{4fchiral} 
\end {eqnarray}

Because of the low energy limit, the bosonic action turns out to 
be bilinear in the bosonic fields. We also observe that the scalar
$\lambda$ has a kinetic term with the opposite sign. This is a
typical phenomenon related with anomalies that has already been
encountered in the 2D quantization of chiral systems. (See for instance
~\cite{Boyanowsky} and next section).
One should expect the stability of the model to be based on some
compensation of states coming from the scalar $\lambda$ with some
scalar states constructed with the tensor $\Lambda$.

\section{Chiral Sch\-win\-ger mo\-del: bosonization}
The Chiral Schwinger model is the simplest model wherein the dynamical
current is always non--conserved.  
 Extended dualization permits to obtain the bosonic 
equivalent of the Chiral Schwinger model 
 including a systematic treatment of the quantization 
arbitrariness. The explicit form of the chiral density current
in terms of two scalar fields is obtained.
\bigskip

The partition function for the Chiral Schwinger model 
is defined by

\begin{equation}
Z_{CSM}(j)=\int{\cal D}a{\cal D}\psi{\cal D}\overline\psi\quad
e^{-S_{CSM}(\psi,\overline\psi,a)-i(j,{\cal J}^{ch})}.\label{zcsm}
\end{equation}
The action has the form
\begin{equation}
S_{CSM}(\psi,\overline\psi,a)={1\over 4}(da,da)
-\overline\psi\gamma^\mu\left[\partial_\mu-ie{1-\gamma_5\over 2}
a_\mu\right]\psi.\label{scsm}
\end{equation}
Here $a^{(1)}$ stands for an abelian gauge field and the chiral
current is
\begin{equation}
{\cal J}^{ch}_\mu= 
\overline\psi\gamma_\mu {1-\gamma_5\over 2}\psi.\label{chcur}
\end{equation}

\medskip

As a result of integrating out the fermion fields we get an effective
action
\begin{equation}
\Gamma_{CSM}(a)=-\log\int{\cal D}\psi{\cal D}\overline\psi\quad
e^{\overline\psi\gamma^\mu\left[\partial_\mu-ie{1-\gamma_5\over 2}    
a_\mu\right]\psi}.\label{gammacsm} 
\end{equation}

\noindent Evaluating this effective action, in its most general
form, one finds~\cite{Jackiw,Harada}
\begin{equation}
\Gamma_{CSM}(a)={e^2\over 2\pi}\left[ \xi(a,a)-
\left(a^+_\mu,{\partial^\mu\partial^\nu\over \partial^2}a^+_\nu\right)
\right].
\label{effxi} 
\end{equation}
Here $a^+_\mu={1\over 2}(\delta_{\mu\nu}+i\epsilon_{\mu\nu})a_\nu$.
The arbitrariness in the quantization procedure is reflected in
the presence of the local term $(a,a)$, and this arbitrariness is
parameterized by the coefficient $\xi$.

This is the result of conventional bosonization~\cite{Jackiw}.  As
a consequence of integrating out the fermi fields, the gauge field
acquires a mass term, depending on the arbitrary parameter $\xi$.
This effective action can be expressed in a local way by introducing
an additional scalar field.

\medskip

Bosonization via extended dualization goes beyond this conventional
result since it permits one to exhibit the bosonic equivalent
of the chiral current.

We see in (\ref{effxi}) that the effective action is never gauge
invariant for any value of $\xi$. Therefore the chiral current,
${\cal J}^{ch}$, is never conserved for any possible quantization of
the Chiral Schwinger model.  Extended dualization is the only
option.

\bigskip

\noindent The partition function (\ref{zcsm}) can be expressed in
terms of a bosonic action
\begin{equation}
Z_{CSM}(j)=\int{\cal D}a{\cal D}[\Lambda]{\cal D}\lambda\quad
e^{-S_b^{CSM}(\Lambda,\lambda,a,j)}.\label{zcsmb}
\end{equation}
The bosonic action, using the extended dualization prescription,
is given by
\begin{equation}
e^{-S_b^{CSM}(\Lambda,\lambda,a,j)}=
\int{\cal D}A{\cal D}\psi{\cal D}\overline\psi\quad
e^{-S_{CSM}(\psi,\overline\psi,a)
-i(j,{\cal J}^{ch})-i(A,{\cal J}^{ch})
+(A,d\lambda+i\ast d\Lambda)}.\label{sbcsm}
\end{equation}

\medskip

\noindent After integrating over the fermions one gets
\begin{eqnarray}
S_b^{CSM}(\Lambda,\lambda,a,j)&=&
+{1\over 4}(da,da)
+(ea+j,d\lambda+i\!\ast\! d\Lambda)  \label{sbgamma} \\
&&-\log\int{\cal D}A
\quad e^{-\Gamma_{CSM}(A)+(A,d\lambda+i\ast d\Lambda)}.\nonumber
\end{eqnarray}

\noindent The gaussian integral over the auxiliary vector field
$A$ is straightforward and the result for the bosonic action is
\begin{eqnarray}
S_b^{CSM}(\Lambda,\lambda,a,j)&=&
+{1\over 4}(da,da)+(ea+j,d\lambda+i\!\ast\! d\Lambda)\label{sbcshfinal} \\
&&+{\pi(4\xi-1)\over 8e^2 \xi^2}(d\Lambda,d\Lambda)
-{\pi(4\xi+1)\over 8e^2 \xi^2}(d\lambda,d\lambda)-
{\pi\over 4e^2 \xi^2}(d\Lambda,d\lambda).\nonumber
\end{eqnarray}

Introducing canonical variables 
\begin{eqnarray}
\theta&=&{\sqrt{\pi}\over 2e\xi}\left\{
(1-2\xi)\Lambda+(1+2\xi)\lambda\right\},\label{changeteta}\\
\phi&=&{\sqrt{\pi}\over e}(\Lambda-\lambda),\nonumber
\end{eqnarray}
the bosonic action becomes
\begin{eqnarray}
S_b^{CSM}(\theta,\phi,a,j)&=&+{1\over 4}(da,da)-{1\over 2}(d\theta,d\theta)
+{1\over 2}(d\phi,d\phi)\nonumber\\
&&+{e\over 2\sqrt{\pi}}
\left(ea+j,d\left[2\xi(\theta+\phi)-\phi\right]\right)
\label{sbcshfinalnorm}\\ 
&&+{e\over 2\sqrt{\pi}}
\left(ea+j,i\!\ast\! d\left[2\xi(\theta+\phi)+\phi\right]\right).\nonumber
\end{eqnarray}
The bosonic action includes the gauge field $a^{(1)}$, plus two
scalar fields $\theta$ and $\phi$, coupled to this gauge field.
The stability of the original model is based on a compensation 
between states coming from the scalar $\theta$ with some combination
of the scalar $\phi$ and the (non--decoupled) zero component
of the gauge field $a$~\cite{Boyanowsky}.

\noindent The bosonic equivalent for the chiral current is given by
\begin{eqnarray}
i\overline\psi\gamma_\mu {1-\gamma_5\over 2}\psi
&\longleftrightarrow& {e\over 2\sqrt{\pi}}
\partial_\mu\left[2\xi(\theta+\phi)-\phi\right] \label{bjchiral}\\ 
&&+i{e\over 2\sqrt{\pi}}\epsilon_{\mu\nu}\partial^\nu
\left[2\xi(\theta+\phi)+\phi\right].\nonumber
\end {eqnarray}

The importance of this result is that bosonization based on extended
dualization allows one to calculate correlation functions of the
chiral current in the most general quantization of the Chiral 
Schwinger model using (\ref{bjchiral}).\footnote{Recently, 
M. Garousi~\cite{Garousi} 
has presented the dualization of the Chiral Schingwer model using only one
scalar field. It is easy to see that his result corresponds to the
particular case $\xi=0$ in our scheme.}

\section{Conclusions}

We have introduced a constructive determination
of the bosonic equivalent of a given anomalous fermi system, called
extended dualization.  The bosonic action includes both a scalar
field and a rank $(D-2)$ antisymmetric form as fundamental fields.
Our extended duality transformation is a generalization of minimal
dualization. The last one applies only for systems in the presence of a
dynamical conserved quantum charge and the only relevant
bosonic field is the $(D-2)$--form.

We have seen that for a given fermionic system the bosonic counterpart
is not unique.  A large freedom is involved in the dualization
transformation.  One can choose the most convenient dualization
in order to have the most tractable bosonic action and correlation
functions for the relevant operators. In general, one can exploit  
this freedom to get specific properties for the bosonic action.
In this paper we have explored
one of the most simple options that is quite efficient for abelian
bosonization when the fermionic effective action is quadratic.

A very nice property of dualization (either minimal or extended)
is that it permits one to study a wide class of fermion systems by
adding some extra model--related terms to the known bosonic version of 
some much simpler fermionic model.
We have examined the bosonization of $D$--dimensional massive
four--fermion interactions by adding appropriate terms to the bosonization
of the $D$--dimensional free massive Dirac fermion.

We have demostrated the utility of extended dualization by
explicitly exhibiting the bosonization of the particular case
of the chiral four--fermion interaction model (in the
low energy limit). The bosonization of this chiral model requires
the inclusion of an additional scalar field in order to yield the
bosonic equivalent of the chiral current density.

We have also applied the extended dualization procedure to determine
the bosonized version of the Chiral Schwinger model and we have 
obtained the expression of the chiral current, for the most
general quantization of the model, in terms of two
independent scalar fields.

In all the above mentioned chiral models, extended dualization has
permitted us to study easily the consequences of the quantization
arbitrariness in the bosonic version of the system.

The considerable freedom in the extended dualization  prescription
introduced in this paper opens up the possibility of applying this
freedom in order to dualize more complicated fermi systems (with
or without anomalies), such as abelian systems in more than two
dimensions beyond the low energy regime, and non--abelian systems.
Moreover, this freedom offers the possibility of imposing new symmetries
on the dual action.

\bigskip

\bigskip\leftline{{\bf Acknowledgments}}\medskip
We wish to thank F. Falceto for discussions.  This work was partially
supported by the CICYT (proyecto AEN 94--0218).  The work of E.R.
has been supported by a contract with the spanish government under
the program  ``contratos para la incorporaci\'on de doctores y
tecn\'ologos a grupos de investigaci\'on en Espa\~na".

\hfill\eject

\appendix{\bf Appendix A}
\bigskip

\noindent In this appendix we will prove the identity 
\protect\begin{equation}
 1=\int {\cal D}A {\cal D}[\Lambda]{\cal D}\lambda\quad 
 e^{F(A,\psi,\overline\psi,\phi)}e^{(A,d\lambda+i\ast d\Lambda)}.
\label{Aidentity}
\end{equation}
We assume that the fields are sufficiently well--behaved
at the boundary that integration by parts is valid. Then:
\begin{itemize}
\item{}
Integrating out the field $\Lambda$, we have
\begin{equation}
d A=0.\label{intL}
\end{equation}
\item{}
To integrate out the field $\lambda$ we need to perform an analytic
continuation $\lambda\rightarrow i\lambda$, and  (at the end of the
calculation) we need to return to the physical region of interest.
This implies
\begin{equation}
d\!\ast\! A=0.\label{intl}
\end{equation}
\end{itemize}

\noindent Therefore, the result  of integrating out the
bosonic fields $\Lambda$ and $\lambda$ is
\begin{equation}
\int {\cal D}[\Lambda]{\cal D}\lambda\quad 
e^{(A,d\lambda+i\ast d\Lambda)}=
\delta (dA)\delta (d\!\ast\! A),\label{intlamb}
\end{equation}
so that the auxiliary vector field must be a harmonic one--form
$A_h^{(1)}$. That is, $\Delta A_h=0$, where $\Delta$ is the
Laplacian that acts on one--forms in the $D$--dimensional space under
consideration.

\noindent Therefore
\protect\begin{equation}
 \int {\cal D}A {\cal D}[\Lambda]{\cal D}\lambda\quad 
 e^{F(A,\psi,\overline\psi,\phi)}e^{(A,d\lambda+i\ast d\Lambda)}=
 \int {\cal D}A_h e^{F(A_h,\psi,\overline\psi,\phi)}.
\label{new}
\end{equation}

\noindent In simple cases where spacetime has trivial topology,
such as $R^n$ or $S^n$, there are no harmonic one--forms, and
therefore the identity (\ref{Aidentity}) follows, provided only
that 
\begin{equation}
F(A=0,\psi,\overline\psi,\phi)=0.\label{cont}
\end{equation}

In more complicated spacetimes with nontrivial topology, 
a modification of (\ref{Aidentity}) is necessary, taking into account
the space of harmonic one--forms.  
\noindent We generalize
the path--integral representation of the identity in the following
way: we replace the integration over the space of 1--forms $A^{(1)}$
by an integration over the space of orbits
\begin{equation}
 [A]=\{A^\prime: A^\prime-A=A_h;\quad \Delta A_h=0\}.\label{harmorbits}
\end{equation}

\noindent The generalization of (\ref{Aidentity}) is obtained by
taking a quotient over the space of harmonic one--forms, and then
averaging the functional $F$ over all harmonic one--form transformations,
\begin{equation}
 1=\int {\cal D}[A] {\cal D}[\Lambda]{\cal D}\lambda\quad 
 e^{\int {\cal D}A_h F(A+A_h,\psi,\overline\psi,\phi)}
 e^{(A,d\lambda+i\ast d\Lambda)}.
\label{newbis}
\end{equation}

\noindent The proof is as follows: Integrating over the bosonic
fields, we have
\begin{equation}
 \int {\cal D}[A] {\cal D}[\Lambda]{\cal D}\lambda\quad 
 e^{\int {\cal D}A_h F(A+A_h,\psi,\overline\psi,\phi)}
 e^{(A,d\lambda+i\ast d\Lambda)}=
 e^{\int {\cal D}A_h F(A_h,\psi,\overline\psi,\phi)}.
\label{newbisbis}
\end{equation}
where translation invariance for the integration over harmonic
1--forms has been used.

\noindent The identity  (\ref{newbis}) follows, provided
that
\begin{equation}
 \int {\cal D}A_h \; F(A_h,\psi,\overline\psi,\phi)=0.
\label{newtris}
\end{equation}
This is the generalization, in the case of nontrivial harmonic
one--forms, of the previous condition $F(A=0,\psi,\overline\psi,\phi)a=0$.

\medskip
If the functional $F(A,\psi,\overline\psi,\phi)$ is linear in 
the auxiliary vector field $A$, as it is the case for the simplest
choice 
$F(A,\psi,\overline\psi,\phi)=-i(A,{\cal J})$ [in (\ref{fchoice})],
the condition (\ref{newtris}) is trivially satisfied and we have
\begin{eqnarray}
\int {\cal D}A_h \; F(A+A_h,\psi,\overline\psi,\phi)= 
F(A,\psi,\overline\psi,\phi)\int {\cal D}A_h+\label{newquat}\\
+\int {\cal D}A_h \; F(A_h,\psi,\overline\psi,\phi)=
F(A,\psi,\overline\psi,\phi)\nonumber
\end{eqnarray}
Therefore, in this case, the only new component to the dualization for
a nontrivial topology comes from the explicit representation
of the integration over harmonic orbits for the auxiliary vector
field $A^{(1)}$ in (\ref{newbis}).
The simplest example, a two dimensional fermion in a cylinder,
has been done in detail in~\cite{BuQue2d}.

\bigskip \bigskip 

\noindent\appendix{\bf Appendix B}
\bigskip 

\noindent This appendix is devoted to a sketch of the technical
details of the  $D=2$ path integral calculation of
\begin{equation}
 e^{-S^{free}_{b}(m;\Lambda,\lambda)}=
 \int{\cal D}A{\cal D}\psi{\cal D}\overline\psi\quad 
 e^{\overline\psi\gamma^\mu(\partial_\mu-i A_\mu)\psi
 +m\overline\psi\psi 
+(A,d\lambda+i\ast d\Lambda)},\label{Bgammasbfreem} 
\end{equation}
for small fermion mass, $m$. 

\medskip

We will proceed by performing a perturbative expansion in the mass, but
first we rearrange this expression in a more convenient form.

\noindent Consider the following local transformation for the
fermion fields:
\begin{eqnarray}
\psi&\longrightarrow& e^{i(\zeta+\varphi\gamma_5)}\psi,
\label{dectransf}\\ 
\overline\psi&\longrightarrow& 
\overline\psi e^{-i(\zeta-\varphi\gamma_5)}.
\nonumber
\end{eqnarray}
The parameters of this transformation are the two scalar fields,
$\zeta$ and $\varphi$, used to write the auxiliary vector field,
$A=d\zeta+i\!\ast\!  d\varphi$.

\noindent Under such a transformation we have 
\begin{equation}
e^{- S^{free}_{b}(m;\Lambda,\lambda)}=
\int{\cal D}A{\cal D}\psi{\cal D}\overline\psi\quad J(A)\quad
 e^{\overline\psi\gamma^\mu\partial_\mu\psi
 +m\overline\psi e^{-2i\varphi\gamma_5}\psi 
+(A,d\lambda+i\ast d\Lambda)}.\label{Bdecsfree} 
 \end{equation}
The Jacobian $J(A)$ associated with the finite transformation
(\ref{dectransf}) can be read off from the general result (\ref{dasj})
in the particular case $\alpha=\zeta$, $\beta=-\varphi$. One finds
\begin{equation}
J(A)={1\over 2\pi}\left[\xi(d\varphi,d\varphi)
+\eta(d\zeta,d\zeta)\right].\label{Bj}
\end{equation}

\noindent Because the mass term depends only on the field $\varphi$,
the integration over $\zeta$ can be easily done. Apart from trivial
constants we get
 \begin{eqnarray}
 S^{free}_{b}(m;\Lambda,\lambda)
 &=&{\pi\over 2\eta} (d\lambda,d\lambda)\label{Bfreevar}\\   
 &&-\log\int{\cal D}\varphi{\cal D}\psi{\cal D}\overline\psi\quad 
 e^{\overline\psi\gamma^\mu\partial_\mu\psi
 +m\overline\psi e^{-2i\varphi\gamma_5}\psi 
+{\xi\over 2\pi}(d\varphi,d\varphi)-(d\varphi,d\Lambda)}.\nonumber
\end{eqnarray}

\noindent Rescaling, to the canonical variable $\varphi \rightarrow
\sqrt{\pi\over\xi}\varphi + {\pi\over\xi}\Lambda$, we have
\begin{eqnarray}
 S^{free}_{b}(m;\Lambda,\lambda)&=&
 {\pi\over 2\eta} (d\lambda,d\lambda)
 +{\pi\over 2\xi} (d\Lambda,d\Lambda)\label{Bvar}\\   
 &&-\log\int{\cal D}\varphi{\cal D}\psi{\cal D}\overline\psi\quad 
 e^{\overline\psi\gamma^\mu\partial_\mu\psi
 +m\overline\psi e^{-2i
 \left(\sqrt{\pi\over\xi}\varphi+{\pi\over\xi}\Lambda\right)\gamma_5}\psi 
+{1\over 2}(d\varphi,d\varphi)}.\nonumber
\end{eqnarray}

\noindent Therefore
\begin{eqnarray}
 S^{free}_{b}(m;\Lambda,\lambda)&=&
 S^{free}_{b}(m=0;\Lambda,\lambda)\label{Bvar2}\\   
 &&-\log\int{\cal D}\varphi\quad 
 e^{+{1\over 2}(d\varphi,d\varphi)
 -\Gamma_m[\sqrt{\pi\over\xi}\varphi+{\pi\over\xi}\Lambda]},\nonumber
\end{eqnarray}

\noindent where the effective action $\Gamma_m$ is the result of
the fermionic integral
\begin{equation}
 e^{-\Gamma_m[\Theta]}=
 \int{\cal D}\psi{\cal D}\overline\psi\quad  
 e^{\overline\psi\gamma^\mu\partial_\mu\psi
 +m\overline\psi e^{-2i\Theta\gamma_5}\psi}. 
 \label{Bg}
\end{equation}

So far we have done nothing more than rearrange the integral in a
convenient way. In order to perform (\ref{Bg}) we will do 
an expansion in powers of the fermion mass.

\noindent Introducing $\sigma_\pm\equiv\overline\psi {1\over 2}
(1\pm\gamma_5)\psi$, one gets
\begin{eqnarray}
 e^{-\Gamma_m[\Theta]}&=&
 \sum_{i=0}^\infty{m^{2i}\over i!i!}
 \int\left[\prod_{k=1}^i d^2 x_k d^2 y_k\right]
 e^{2i\sum_k[\Theta(y_k)-\Theta(x_k)]}
 \label{Bgm}\\ 
 &&\int{\cal D}\psi{\cal D}\overline\psi\quad  
 e^{\overline\psi\gamma^\mu\partial_\mu\psi}
 \prod_{k=1}^i\sigma_-(x_k) \sigma_+(y_k).\nonumber 
\end{eqnarray}

\noindent Using Zinn--Justin~\cite[p.~680, eq.~(A28.14)]{masspath}, 
one sees
\begin{eqnarray}
 \int{\cal D}\psi{\cal D}\overline\psi  
 e^{\overline\psi\gamma^\mu\partial_\mu\psi}
 \prod_{k=1}^i\sigma_-(x_k) \sigma_+(y_k)&=&
 \left({1\over 2\pi}\right)^{2i}
 {\prod^i_{k<l} |z_k-z_l|^2 |z^\prime_k-z^\prime_l|^2
 \over \prod^i_{k,l} |z_k-z^\prime_l|^2}
 \nonumber\\ 
 &\equiv&K_i(x,y),\label{zinn1}
\end{eqnarray}
where $z_k=x^0_k+ix^1_k$, $z^\prime_k=y^0_k+iy^1_k$.
		  
\medskip

\noindent This implies
\begin{eqnarray}
 S^{free}_{b}&=&
 {\pi\over 2\eta} (d\lambda,d\lambda)+{\pi\over 2\xi} 
 (d\Lambda,d\Lambda)\label{Bvar3}\\   
 &&-\log\sum_{i=0}^\infty{m^{2i}\over i!i!}
 \int\left[\prod_{k=1}^i d^2 x_k d^2 y_k\right] K_i(x,y)
e^{i{2\pi\over\xi}\sum_k[\Lambda(y_k)-\Lambda(x_k)]}\nonumber\\ 
&&\int{\cal D}\varphi\quad 
 e^{+{1\over 2}(d\varphi,d\varphi)}
\prod_{k=1}^i e^{2i\sqrt{\pi\over\xi}\varphi(y_k)}
\prod_{k=1}^i e^{-2i\sqrt{\pi\over\xi}\varphi(x_k)}.\nonumber
\end{eqnarray}

\noindent Now using Zinn--Justin~\cite[p.~664, eq.~(28.13)]{masspath}, 
\begin{equation}
\int{\cal D}\theta\quad 
 e^{-{1\over 2t}(d\theta,d\theta)+i\sum_i\epsilon_i\theta(x_i)}
 \propto\left\{\begin{array}{ll}0&\mbox {for $\sum_i\epsilon_i\neq 0$}\\
\prod_{i<j}
\left(\Lambda_{UV}|x_i-x_j|\right)^{\epsilon_i\epsilon_j t\over 2\pi}&
\mbox {for $\sum_i\epsilon_i= 0$}\end{array}\right.\label{zinnc2}
\end{equation}
for an ultraviolet cutoff $\Lambda_{UV}$ that appears when one regularizes
the free boson propagator.  The fact that this
correlation function is zero unless the coefficients satisfy the
condition $\sum_i\epsilon_i= 0$ is a result of invariance under
constant translations of the field $\theta$.

\noindent The integral over the scalar fields $\varphi$ results in
\begin{equation}
\int{\cal D}\varphi\quad 
 e^{+{1\over 2}(d\varphi,d\varphi)}\quad
\prod_{k=1}^i e^{2i\sqrt{\pi\over\xi}\varphi(y_k)}
\prod_{k=1}^i e^{-2i\sqrt{\pi\over\xi}\varphi(x_k)}
\propto(\Lambda_{UV})^{2i}K^{-1}_i(x,y).\label{zinn2}
\end{equation}

\noindent The final result is
\begin{eqnarray}
 S^{free}_{b}(m;\Lambda,\lambda)&=&
 {\pi\over 2\eta} (d\lambda,d\lambda)
 +{\pi\over 2\xi} (d\Lambda,d\Lambda)\label{Bsbfinal}\\   
 &&-\log\sum_{i=0}^\infty{(m\Lambda_{UV})^{2i}\over 2i!}
 \left(\begin{array}{c}2i\\i\end{array}\right)
 \left(\int d^2 x e^{i{2\pi\over\xi}\Lambda}\right)^i
 \left(\int d^2 x e^{-i{2\pi\over\xi}\Lambda}\right)^i.\nonumber
\end{eqnarray}

\noindent Consider now the partition function in the presence of
a source $j$:
\begin{equation}
Z=\int{\cal D}[\Lambda]{\cal D}\lambda\quad 
e^{-S^{free}_{b}(m;\Lambda,\lambda)
-(j,d\lambda+i\ast d\Lambda)}.\label{zagain}
\end{equation}

\noindent Invoking invariance under constant translations of the
field $\Lambda$,  we can show that the result of (\ref{zinnc2}) for
$\sum_i\epsilon_i\neq 0$ generalizes, in the presence of the source
$j$, to
\begin{eqnarray}
Z &=
  \int{\cal D}[\Lambda]{\cal D}\lambda&\quad 
  e^{-{\pi\over 2\eta} (d\lambda,d\lambda)-
  {\pi\over 2\xi} (d\Lambda,d\Lambda)-(j,d\lambda+i\ast d\Lambda)}
\label{proof}\\
&&\sum_{i=0}^\infty{(m\Lambda_{UV})^{2i}\over 2i!}    
  \left(\begin{array}{c}2i\\i\end{array}\right)
  \left(\int d^2 x e^{i{2\pi\over\xi}\Lambda}\right)^i
  \left(\int d^2 x e^{-i{2\pi\over\xi}\Lambda}\right)^i 
\nonumber\\
&=
   \int{\cal D}[\Lambda]{\cal D}\lambda&\quad 
   e^{-{\pi\over 2\eta} (d\lambda,d\lambda)-
   {\pi\over 2\xi} (d\Lambda,d\Lambda)-(j,d\lambda+i\ast d\Lambda)}
\nonumber\\
&& \sum_{i=0}^\infty{(m\Lambda_{UV})^{2i}\over 2i!}     
   \sum_{k=0}^{2i}\left(\begin{array}{c}2i\\k\end{array}\right)
   \left(\int d^2 x e^{i{2\pi\over\xi}\Lambda}\right)^k
   \left(\int d^2 x e^{-i{2\pi\over\xi}\Lambda}\right)^{2i-k}
\nonumber\\
&=
  \int{\cal D}[\Lambda]{\cal D}\lambda&\quad 
  e^{-{\pi\over 2\eta} (d\lambda,d\lambda)-
  {\pi\over 2\xi} (d\Lambda,d\Lambda)-(j,d\lambda+i\ast d\Lambda)}
\nonumber\\
&&\sum_{i=0}^\infty{(m\Lambda_{UV})^{2i}\over 2i!}      
 \left(\int d^2 x e^{i{2\pi\over\xi}\Lambda}+
 \int d^2 x e^{-i{2\pi\over\xi}\Lambda}\right)^{2i}
\nonumber\\  
&=
  \int{\cal D}[\Lambda]{\cal D}\lambda&\quad 
  e^{-{\pi\over 2\eta} (d\lambda,d\lambda)-
  {\pi\over 2\xi} (d\Lambda,d\Lambda)-(j,d\lambda+i\ast d\Lambda)}
\nonumber\\
&&\sum_{i=0}^\infty{(m\Lambda_{UV})^{i}\over i!}      
 \left(\int d^2 x e^{i{2\pi\over\xi}\Lambda}+
 \int d^2 x e^{-i{2\pi\over\xi}\Lambda}\right)^{i}
\nonumber\\  
&=
  \int{\cal D}[\Lambda]{\cal D}\lambda&\quad 
  e^{-{\pi\over 2\eta} (d\lambda,d\lambda)-
  {\pi\over 2\xi} (d\Lambda,d\Lambda)-(j,d\lambda+i\ast d\Lambda)}
\nonumber\\
&&\exp\left({2m\Lambda_{UV}\int d^2 x \cos{\left[{2\pi\over\xi}
\Lambda\right]}}\right)
\nonumber  
\end{eqnarray}

\noindent This result is a generalization of the well 
known~\cite{Coleman,Mandelstam}
equivalence of a massive Dirac fermion in $D=2$ to
the sine--Gordon theory, for the particular value $\xi=1$.

\vfill
\eject

\newpage

\end{document}